ssh 
\documentclass[aps,prl,twocolumn,groupedaddress,floatfix,showpacs]{revtex4}

\usepackage{subfigure}
\usepackage{color}
\usepackage{graphicx}
\usepackage{amssymb}
\usepackage{amsmath}
\usepackage{natbib}

\begin{document}

\newcommand{\ii}{\mathrm{i}}
\newcommand{\kk}{\mathbf{k}}
\newcommand{\rr}{\mathbf{r}}
\newcommand{\qq}{\mathbf{q}}

\parskip=0pt

\title{Coexistence of ferromagnetism and superconductivity near 
quantum phase transition: The Heisenberg- to Ising-type crossover}



\author{Andriy H. Nevidomskyy}

\email[]{ahn22@phy.cam.ac.uk}

\affiliation{Theory of Condensed Matter, Cavendish Laboratory, University
of Cambridge, Cambridge CB3 0HE, UK}


\date{\today}

\begin{abstract}
A microscopic mean-field theory of the phase coexistence between
ferromagnetism and superconductivity in the
weakly ferromagnetic itinerant electron system is constructed, while
incorporating a realistic 
mechanism for superconducting pairing due to the exchange of critical
spin fluctuations. The self-consistent solution of the resulting
equations determines the superconducting transition temperature which
is shown to depend strongly on the exchange splitting.
The effect of phase crossover from isotropic (Heisenberg-like) to
uniaxial (Ising-like) spin fluctuations near the quantum phase
transition is analysed and the generic phase diagram is obtained.  
This scenario is then applied to the case of
itinerant ferromagnet ZrZn$_2$, which sheds light on the
proposed phase diagram of this compound. Possible explanation of
superconductivity in UGe$_2$ is also discussed.
\end{abstract}


\maketitle

There has been extensive experimental research done recently on the possible
coexistence of superconductivity (SC) and ferromagnetism (FM) in
strongly correlated electron materials. 
First, superconductivity was discovered in the ferromagnetic
metal UGe$_2$ at high pressure\cite{Saxena00}. 
Later a low-temperature SC phase was found in another f-electron
compound URhGe\cite{Aoki01} and in the d-electron ferromagnet ZrZn$_2$
\cite{Pfleiderer01_ZrZn2}.
In the best studied cases of UGe$_2$ and URhGe the experiment
strongly suggests that superconductivity coexists with itinerant
ferromagnetism in these compounds. 
Notably, it is the same electrons that are involved in both SC and FM,
which leads to inter-dependence of the corresponding order parameters.

The aim of this paper is to construct a mean-field theory of
the phase coexistence between FM and SC on the border of magnetism,
while adopting a realistic mechanism for superconducting pairing.
We show that interplay between FM and SC order parameters
has crucial effect on the resulting phase diagram. 
We then propose a mechanism that would explain the enhancement of SC
transition temperature in the FM phase 
and discuss
the application of this mechanism to 
ZrZn$_2$ and UGe$_2$. 


The fact that SC is observed inside the FM region imposes strict
limitations on the nature of the SC state. 
The very large internal molecular field due to the exchange
interaction (measured \cite{Tsutsui99} to be $\sim 240$~T in UGe$_2$) excludes, due to
the Pauli limitation, not only any singlet-pairing SC but also any
unitary triplet states \cite{Machida01}.
In this paper we analyse consequences of the so-called \emph{non-unitary}
triplet SC state \cite{Machida01,Mineev_book}
on the resulting phase diagram, and then develop a microscopic
theory based on spin-fluctuation mediated pairing
to proceed beyond the phenomenological level of
treatment reported in Ref. \cite{Machida01}.



%

A nonunitary triplet state is described \cite{Machida01} by the order parameter 
$\hat{\Delta}_{\alpha\beta}(\mathbf{k})\equiv \langle c_{\mathbf{k},\alpha} c_{\mathbf{k},\beta}\rangle = [i(\mathbf{d(k}) \cdot \mbox{\boldmath$\sigma$})\sigma_y]_{\alpha\beta}$, 
where {\boldmath $\sigma$} $=\hat{\mathbf{x}}\sigma_x +\hat{\mathbf{y}}\sigma_y +
\hat{\mathbf{z}}\sigma_z $ denote the usual Pauli matrices and the
basis of symmetric matrices $i\mbox{\boldmath$\sigma$}\sigma_y$ was
used to represent odd angular momentum pairing. The
three-dimensional complex vector $d(\mathbf{k})$ fully characterizes
the triplet pairing state \cite{Mineev_book}. 
In what follows we assume for simplicity an easy axis of
magnetization in the $z$-direction.
Because of the pair-breaking effect of strong exchange field $M$, only
the Cooper pairs with parallel spins will survive. In this case of
equal-spin pairing we can write vector ${\bf d}$ in the form 
${\bf d}=(d_x,d_y,0)$. Denoting $\Delta_\pm \equiv d_x \pm id_y$, the SC order parameter becomes
\begin{equation}
\hat{\Delta}(\mathbf{k})_{\alpha \beta}=
\left( \begin{array}{ccc}
-\Delta_-(\mathbf{k}) & 0 \\
                    0 & \Delta_+(\mathbf{k})
\end{array}\right).
\end{equation}

We shall start from the effective Heisenberg model for itinerant
electrons with spin 
$\mathbf{s}(\rr)=\sum_{\alpha\beta} \psi^\dagger_\alpha(\rr)
\boldsymbol{\sigma}_{\alpha\beta}\psi_\beta(\rr)$, where
$\psi^\dagger_\alpha(\rr)$, $\psi_\alpha(\rr)$ are electron field
operators.
Some attractive pair-forming interaction $\hat{V}$ is also assumed:
\begin{eqnarray}
H_{FM+SC}  =  \sum_{\mathbf{k,\alpha}}\epsilon_\mathbf{k}\, c^\dagger_{\mathbf{k},\alpha}
c_{\mathbf{k},\alpha} 
- I \int d\mathbf{r}\: \mathbf{s}(\mathbf{r}) \cdot \mathbf{s}(\mathbf{r})
                                                 \nonumber\\
 +  \frac{1}{2} \sum_{{\bf k},{\bf k'}} V_{\alpha\beta,\lambda\mu}
						   ({\bf k},{\bf k'})
\, c^\dagger_{-{\bf k}\alpha}\, c^\dagger_{{\bf k}\beta}\, c_{{\bf k'}\lambda}\, c_{-{\bf k'}\mu}.
\end{eqnarray}
Making use of the Hubbard--Stratonovich transformation and
 integrating out fermionic degrees of freedom in order to
 arrive at the effective action 
in terms of the bosonic
 field operators $\hat{\Delta}_\pm (\mathbf{k})$ and $M(\rr)\equiv s_z(\rr)$, we 
then
 deduce the mean-field equations for the 
order parameters in the \emph{saddle-point} approximation. 
The resulting equations for the SC order parameter have 
usual BCS-like form:
\begin{equation}
\left\{ \begin{array}{l}
\Delta_{-}({\bf k}) = -\frac{1}{V}\sum\limits_{\bf k'} V({\bf k},{\bf
  k'}) \frac{1-2f(E_{-}({\bf k'}))}{2E_{-}({\bf k'})} 
\Delta_{-}({\bf  k'})\\
\Delta_{+}({\bf k}) = -\frac{1}{V}\sum\limits_{\bf k'} V({\bf k},{\bf
  k'}) \frac{1-2f(E_{+}({\bf k'}))}{2E_{+}({\bf k'})} 
\Delta_{+}({\bf  k'}) 
\end{array}  \right. , 
\label{BCS-eqn}
\end{equation} 
where $f(E)$ is the Fermi-Dirac distribution function. 
 
The magnetic order parameter $M$ enters above equations via the
quasiparticle spectrum 
$E_{\pm}({\bf k})=\sqrt{(\epsilon_{\bf k}\mp M)^2 + 
|\Delta_{\pm}({\bf k})|^2}$.
The equation for $M$ looks as
\begin{eqnarray}
\frac{2M}{I}=\frac{1}{V} \sum\limits_{\bf k} \left\{  
\frac{\epsilon_{\bf k}^{\uparrow}\,[1-2f(E_{-})]} {2E_{-}({\bf k})} 
-\frac{\epsilon_{\bf k}^{\downarrow}\,[1-2f(E_{+})]} {2E_{+}({\bf k})}
\right\}
\label{FM-eqn}
\end{eqnarray} 
where $\epsilon_{\bf k}^{\uparrow, \downarrow}\equiv {\epsilon_{\bf
    k}\pm M}$. 
In the limit of pure magnetism (i.e. $\Delta_{\pm}\to 0$ when $T\to 0$), 
this equation reduces to the Stoner criterion for itinerant ferromagnet
$1/I=1/V \sum_{\bf k} \left( -\frac{\partial f}{\partial \epsilon_{\bf
    k}} \right) \approx  N(0)$. 

\begin{figure}[!hbt]
\begin{center}
\includegraphics[width=8.6cm]{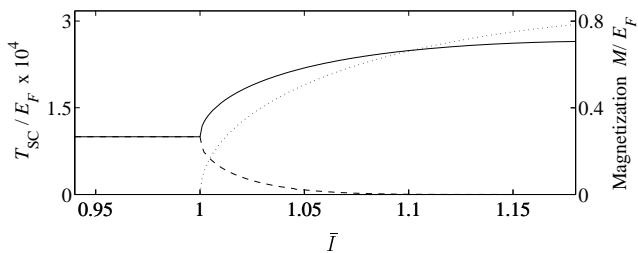}
\end{center}
\caption{\label{Fig.interplay} 
The calculated SC transition temperature as a function of interaction
strength $\bar{I}$ obtained as a result of
solving self-consistent Eqs. (\ref{BCS-eqn},\ref{FM-eqn}) in the weak-coupling BCS
approximation $T_{\mathrm{SC}}\approxeq 1.14\, \omega_c\, 
\exp(-1/\lambda_1^{\sigma})$, where frequency cutoff $\omega_c$ 
was chosen arbitrarily $\omega_c=0.01\,E_F$.
The solid (dashed) line shows $T_{SC}$ for majority
(minority) spin in units of $E_F$. 
Exchange splitting $M$ is plotted with dotted line (right scale). 
}
\end{figure}

In order to illustrate the interplay between the FM and SC order
parameters that follows from this model, we solved
Eqs. (\ref{BCS-eqn},\ref{FM-eqn}) self-consistently for the simple case
of spherical Fermi surface at half filling, while assuming that SC
pairing strength in the p-channel $V(\mathbf{k},\mathbf{k'})\equiv
V_{l=1}(k,k')\sum\limits_{m=-1}^{1} Y_{1m}(\mathbf{\hat{k}})
Y_{1m}^{*}(\mathbf{\hat{k'}})$ has BCS-like form (i.e. $V_{1}(k,k')$
vanishes everywhere except the narrow region near the Fermi surface)
and does not depend on exchange interaction $I$. 
The resulting SC transition temperature was calculated in the weak-coupling BCS
approximation and is shown in Fig. \ref{Fig.interplay} as a
function of dimensionless interaction constant $\bar{I}\equiv N(0)I$. 
It is apparent that exchange splitting has large effect on
superconductivity, enhancing $T_{\mathrm{SC}}$ in the FM phase for the
majority spin channel and suppressing it for the minority spin. 
This is not surprising since
exchange splitting enhances (suppresses) the density of states (DOS)
$N_\sigma$ in the majority (minority) 
spin channel, which enters the expression for the dimensionless
pairing strength $\lambda_1^{\sigma}=N_\sigma(0) |V_{1}|$.
We note that at this stage the symmetry of the 
magnetic state has no effect on $T_{\mathrm{SC}}$ since the mass
renormalization effects have not been taken into account when
calculating $T_{\mathrm{SC}}$. These effects will prove to be very
important in what follows.

We shall now address the issue of the mechanism of the superconducting
pairing that arises in the vicinity of the magnetic quantum phase transition.
It has been shown that exchange of spin fluctuations, called paramagnons,
can lead to an attractive pairing interaction. On the paramagnetic
side the strength of this interaction was derived in the context of
superfluidity in $^3$He \cite{Berk_Schrieffer66, Anderson_Brinkman73,
  Nakajima73}. 
On the FM side of the transition the corresponding formulae were obtained 
in Ref. \cite{Brinkman_Engelsberg68, Fay_Appel80}.
Following their approach, the (attractive) equal-spin
pairing interaction is given by  
\begin{equation}
V^{\sigma\sigma}(k,k+q)= - \left(
\frac{I^2\chi_0^{-\sigma}} {1-I^2\chi_0^\sigma \chi_0^{-\sigma}}
\right)_q \qquad (\sigma=\uparrow, \downarrow),
\label{V(q)}
\end{equation}
where $\chi_0^\sigma(q) \equiv \chi_0^{\sigma \sigma}(i\omega_l,{\bf
  q})$ is 
the Lindhard function 
of the non-interacting system in the given spin channel.
The usual BCS pairing parameter $\lambda$ in the triplet channel
is given by $\lambda_{l=1}^\sigma \equiv
N_\sigma(0)|V_1^{\sigma\sigma}|$, where $N_\sigma(0)$ is the DOS at
Fermi level and 
$V_l^{\sigma\sigma}$ is the strength of the interaction  in the
$l$-orbital channel, which for spherical Fermi surface is
\begin{eqnarray}
V_l^{\sigma\sigma}=\int_0^{2 k_{F\sigma}} 
\frac{q \mathrm{d}q} {2 k_{F\sigma}^2} 
P_l\left( 1-\frac{q^2}{2 k_{F\sigma}^2} \right) 
                            \nonumber \\
\times V^{\sigma\sigma}(\omega=0,q)
\bigg\vert_{|{\bf k}|=|{\bf k}+{\bf q}|= k_{F\sigma}} ,
\label{lambda}
\end{eqnarray}
where $P_l(z)$ denotes the Legendre polynomial of order $l$.
The dominant contribution to the integral comes from the small-$q$
region since $V^{\sigma\sigma}(q)$ is strongly peaked for
$q\to 0$. This allows us to employ the small-$q$ approximation for the
Lindhard function $\chi_0^{\sigma}({\bf q})
\approxeq N_\sigma(0)[1-\frac{1}{12}(q/k_{F\sigma})^2 + O(q^4)]$, which 
enters Eq. \ref{V(q)}.


The mass enhancement near magnetic instability renormalizes
the BCS pairing parameter $\lambda_1$ to the new value
$\lambda^{\sigma*}=\lambda_1^{\sigma}/Z_\sigma(0)\equiv
\lambda_1^{\sigma}/(1+\lambda_0^\sigma)$,
where $\lambda_0$ is the $s$-wave pairing interaction parameter and 
$Z(0)$ is the mass enhancement factor\footnote{Here we defined
  $Z(0)\equiv m^*/m$
  as an inverse of the quasiparticle renormalization constant
  $z_{k_F}=1/Z(0)$, following the notation of
  Refs. \cite{Fay_Appel80,Levin_Valls78}} at the Fermi surface given by 
\begin{equation}
\frac{m^*}{m} \approxeq Z(0) = 
\left( 1-\frac{\partial\Sigma(\omega,|{\bf k}|=k_F)} {\partial\omega} 
\right) \bigg\vert_{\omega=0},
\end{equation}
where $\Sigma$ is the single-particle self-energy. 

\begin{figure}[tbh]
\begin{center}
\includegraphics[width=8.0cm]{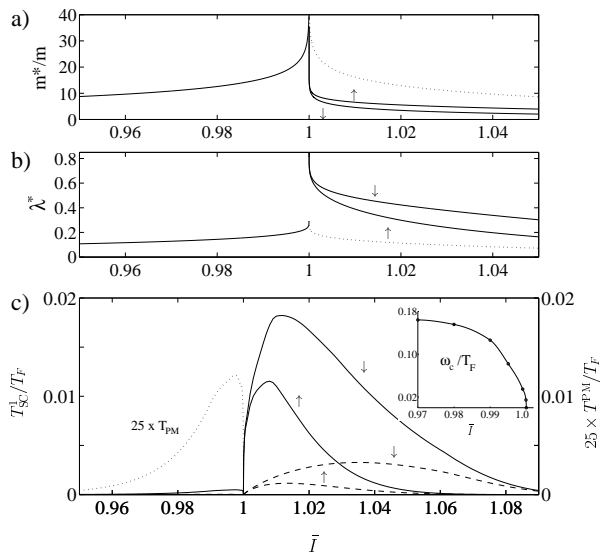}
\end{center}
\caption{\label{mass_pairing_TSC} 
a), b): Calculated properties as functions of the exchange splitting
$\bar{I}$ of the itinerant FM with proposed
Heisenberg- to Ising-type crossover occurring at $\bar{I}=1$.
Spin fluctuations of Heisenberg type are assumed in the PM phase and
of Ising-type in the FM phase (solid line). 
This is to be compared with purely Heisenberg-like behaviour
on both sides (dotted line).
a) Mass enhancement parameter $Z_\sigma(0)$, where formulae of
Ref. \cite{Brinkman_Engelsberg68} were adopted for 
calculating the mass enhancement contributions $\lambda_0^{\sigma,L}$
and $\lambda_0^{\sigma,T}$. 
b) The effective pairing parameter
$\lambda^{\sigma*}=\lambda^{1\sigma}/Z_\sigma(0)$.
c) The calculated superconducting $T_\mathrm{SC}$ (solid curve, left
scale). 
The frequency cutoff $\omega_c$, shown in the inset,
was extracted from the strong-coupling calculation 
by Levin and Valls \cite{Levin_Valls78} and is symmetric about
$\bar{I}=1$ point ($T_F$ is the Fermi temperature). 
The broken curve (left scale) shows $T_\mathrm{SC}$ if one adopts the 
phenomenological formula for $\omega_c$ from Ref.~\cite{Fay_Appel80}, 
and the dotted curve (right scale) shows
$T_\mathrm{SC}$ in the PM phase scaled by a factor of 25. 
}
\end{figure}
It is remarkable that experimentally no SC is observed in the paramagnetic (PM)
phase of UGe$_2$ and ZrZn$_2$. Admittedly, this is
different from the result of Fay and Appel \cite{Fay_Appel80} who
obtained comparable values of $T_\mathrm{SC}$ on both sides of the
magnetic transition.
 Here we propose a tentative generic explanation for the strong
 suppression of SC on the PM side of the transition 
in the nearly ferromagnetic metal, which is due to the
different nature of spin fluctuations on the two sides of the magnetic
quantum phase transition. 
We consider a nearly 
ferromagnetic 
 metal which has no
preferred magnetization axis in the PM phase and thus is characterized
by spin fluctuations that 
are of Heisenberg type.
In this case both longitudinal and transverse spin fluctuations
contribute to the effective mass enhancement, so that
$Z(0)=1+\lambda_0^{L}+\lambda_0^{T}$.
Consequently, the renormalized pairing strength 
$\lambda^* = \lambda_1/Z(0)$ and the SC transition temperature are both
small on the PM side. Remarkably, higher values of $T_{\mathrm{SC}}$ on
the FM side could be achieved if spin fluctuations there were of Ising-type,
so that only longitudinal spin fluctuations contribute to the effective
mass enhancement $Z(0)=1+\lambda_0^{L}$.
 This would lead to $Z(0)$
being about three times larger in the PM phase than it is in the FM
phase, as illustrated in Fig.~\ref{mass_pairing_TSC}a). The resulting
pairing parameter $\lambda^{\sigma*}$ is shown in
Fig.~\ref{mass_pairing_TSC}b). We will first outline the
consequences of this hypothesis on $T_{\mathrm{SC}}$, and then turn to the discussion of its
validity.

The SC transition temperature is notoriously
difficult to calculate. For the purpose of comparison of $T_{\mathrm{SC}}$ on
both sides of the FM transition a simple McMillan-type formula \cite{McMillan68}
should suffice:
\begin{equation}
T_\mathrm{SC}^{1,\sigma} \approxeq 1.14\, \omega_c \,
\exp[-1/ \lambda_1^{\sigma*}],
\label{Eq_Tsc}
\end{equation}
where cutoff $\omega_c$ simulates in a crude way the fact that in reality
$V^{\sigma\sigma}(\omega,\mathbf{q})$ is strongly frequency-dependent, being sharply
peaked at small energy transfers.
It turns out \cite{Levin_Valls78} that $\omega_c$ depends strongly on exchange interaction,
as shown in the inset of Fig. \ref{mass_pairing_TSC}c).
The resulting dependence
$T_{SC}(\bar{I})$ is plotted in Fig.~\ref{mass_pairing_TSC}c), 
which indicates clearly that the SC
transition temperature is an order of magnitude higher in the FM
phase than it is in the PM phase.

Our calculations
suggest that $T_\mathrm{SC}$ goes through a maximum  
and then approaches zero at the quantum critical point, in accordance
with Ref. \cite{Fay_Appel80, Levin_Valls78}.
A recent strong-coupling calculation by Roussev and Millis
\cite{Roussev_Millis01} suggests however that $T_\mathrm{SC}>0$ generically at
the magnetic critical point, contrary to our result.
We note that though interesting from the
fundamental point of view, the behaviour of $T_\mathrm{SC}$ directly
at the magnetic phase transition is not so important in practice, since
experimentally magnetic transition proves to be first order
\cite{Saxena00,Aoki01}, thereby eliminating the low 
values of $(\bar{I}-1)$ from consideration.  




For outlined scenario to take place, two crucial
conditions are necessary. 
Firstly, the contribution of soft transverse spin fluctuations
to $\lambda_0$ must be quenched on the FM side. 
This is achieved due to spin waves taking over the available phase
space as the magnetization $M$ increases. 
Indeed, the
fraction of the momentum space available to gapless spin fluctuations
is  $q\geq q_\perp \equiv k_{F\downarrow}- k_{F\uparrow}$, the rest being taken by
spin waves at $0\leq q\leq q_\perp$. Long-wavelength spin waves themselves
do not contribute to the mass enhancement in the leading order of the
perturbation theory \cite{Izuyama64}. Thus as the exchange splitting
increases, 
the soft spin fluctuations shift to larger $q$-values, 
thereby decreasing their contribution to $\lambda_0$. 

However, this suppression also affects
the pairing strength $\lambda_1 \propto \chi_\Vert$ due to
longitudinal spin fluctuations which become quenched as well. 
The situation can be cured by the second condition: the existence of the
quantum meta-magnetic transition (MMT) \cite{Millis02} somewhere in
the FM phase. Indeed, the longitudinal susceptibility $\chi_\Vert$ is
peaked near the jump in magnetization accompanying such a
transition, as seen experimentally in UGe$_2$ \cite{Huxley03} 
and in Sr$_3$Ru$_2$O$_7$ \cite{Perry04}. 
As a result, longitudinal spin fluctuations will be
enhanced and the material 
will appear effectively Ising-like near the MMT, justifying the
assumption made above.
However unlikely the ``coincidental'' presence of meta-magnetic
transition near the quantum transition to the FM phase may appear
at first sight, the experiment suggests that 
this is not uncommon in the ferromagnetic strongly correlated electron
materials. 
Indeed, the MMT has been observed in
Sr$_3$Ru$_2$O$_7$ \cite{Grigera01}, UGe$_2$ \cite{Pfleiderer02_UGe2,
  Huxley03}, and recently in ZrZn$_2$ \cite{Kimura04}.



\begin{figure}[!tb]
\begin{center}
\includegraphics[width=5.8cm]{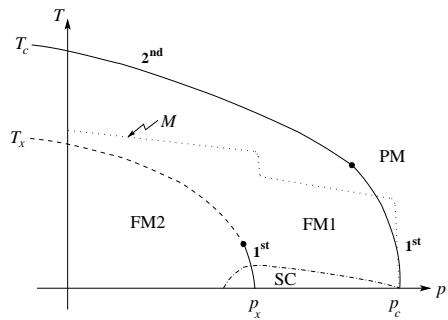}
\end{center}
\caption{\label{UGe2_phase_diag} 
The schematic $p$-$T$ phase diagram of a generic itinerant ferromagnet that
exhibits a meta-magnetic transition at $p=p_x$, associated with the
SC region inside the FM phase, based on studies of ZrZn$_2$ and UGe$_2$.
Pressure $p_x$ is negative in ZrZn$_2$ 
and positive in UGe$_2$. 
The dependence of low temperature magnetization, $M$, on pressure is
shown schematically with dotted line \cite{Tateiwa01,Uhlarz04}.
$T_c$ is the Curie temperature; $T_x$ is a crossover line between
two different ferromagnetic phases, FM1 and FM2.
The filled circles denote critical endpoints, below which the
transitions are first order. 
The dashed line indicates a crossover, rather than a sharp phase transition.
}
\end{figure}

We now turn to the application of the above model to the
experimentally studied materials.
Figure \ref{UGe2_phase_diag} shows the generic phase diagram of an
itinerant ferromagnet that arises from studies of ZrZn$_2$
\cite{Pfleiderer01_ZrZn2,Kimura04,Uhlarz04} and UGe$_2$
\cite{Saxena00,Huxley01,Pfleiderer02_UGe2,Sandeman03}.  
The Curie temperature $T_c$ is suppressed to zero at pressure $p_c$,
where the transition appears to be first order
\cite{Pfleiderer01_ZrZn2,Saxena00} in both compounds.
Another feature, the crossover line $T_x$ between the two
ferromagnetic phases, FM1 and FM2, is also shown. This crossover exhibits itself as
an anomaly in the measurements of resistivity \cite{Oomi98,Saxena00,Tateiwa01} and specific heat \cite{Tateiwa01,Tateiwa04} in UGe$_2$, where it
occurs at pressure $p_x\approx 12$~kbar. In the case of
ZrZn$_2$ $p_x$ appears to be negative \cite{Kimura04, Uhlarz04} 
and thus cannot be probed directly, however
the MMT can be rendered to positive pressures by
applying external magnetic field, where it
has been studied in the de Haas--van Alphen \cite{Kimura04}
and magnetization measurements \cite{Uhlarz04}.

It is notable that SC is observed only on the FM side of the
transition in ZrZn$_2$ \cite{Pfleiderer01_ZrZn2}, but not in the PM
phase. 
This fact can be readily explained by the scenario
proposed in this work. Indeed, ZrZn$_2$ is a three-dimensional
ferromagnet with cubic symmetry, from which the isotropic (Heisenberg)
nature of spin fluctuations in the PM phase is deduced. By contrast, it
follows from the proposed model that spin
fluctuations develop Ising-like symmetry upon entering the FM
phase, where the existence \cite {Kimura04, Uhlarz04} of meta-magnetic
transition at $p=p_x$ is crucial to our argument. 
It thus follows from our calculations
that the SC transition
temperature must be strongly suppressed in the PM phase of ZrZn$_2$.

The situation is more intricate in UGe$_2$, where evidence of strong
uniaxial anisotropy exists on both sides of magnetic transition at
$p_c$ \cite{Saxena00, Huxley03}. However in the light of recent measurements
of specific heat \cite{Tateiwa04} it becomes evident that the very narrow SC
region is centred around the MMT at $p_x$ rather than $p_c$. This can be easily
understood given that the transition at $p_c$ is strongly first order
\cite{Saxena00} and therefore has a strong pair-breaking effect on
SC. It is hence not surprising that no SC is seen on both sides of
$p_c$. The presence of SC in UGe$_2$ is instead due to critical spin fluctuations at
$p_x$, which is only a weakly first order transition. We note that
this view has been expressed already in earlier works on UGe$_2$
\cite{Huxley01, Tateiwa04}.
Our proposed theoretical model has thus an indirect application to UGe$_2$
in a sense that the observed uniaxial (Ising) symmetry enhances
$T_\mathrm{SC}$, which would have been much more suppressed if the
Heisenberg-like spin fluctuations had prevailed in this compound.


The above argument already suggests that SC phase must be suppressed
both in FM1 and PM phases of UGe$_2$ close to $p_c$.
It should be noted that the qualitative change in the Fermi surface
observed at the magnetic transition by de Haas--van Alphen experiment
\cite{Terashima01} may be another factor that suppresses SC near $p_c$.
In this context, the existence of a double peak structure in the
electronic density of states very close to the Fermi level has been
proposed \cite{Sandeman03} as a possible microscopic explanation.
We also note that absence of SC in the PM phase of UGe$_2$ may be partly
due to the spin degeneracy of the Fermi surface 
as it can, in principle, enhance spin-flip processes of the electrons
forming a Cooper pair, which would have detrimental consequence on
spin-triplet SC state. 


In conclusion, we have formulated a mean-field theory of coexisting FM and
SC in terms of the equations for the corresponding order parameters
that have to be solved self-consistently.  
We have also incorporated a microscopic mechanism of the SC pairing due to the
 exchange of spin fluctuations in our model. 
A scenario based on Heisenberg- to Ising-type crossover has been proposed,
which provides a natural explanation of the enhancement of SC on the
FM side of magnetic transition, observed experimentally in ZrZn$_2$.  
The apparent suppression of SC in the PM phase of UGe$_2$ is
explained by the detrimental effect that the strongly first-order
phase transition at $p_c$ has on pair-forming spin fluctuations.  
The proposed theoretical model supports the evidence of SC
in UGe$_2$, as superconductivity is predicted to be enhanced by Ising-like 
spin fluctuations near $p_x$.

The author would like to thank Gilbert Lonzarich, Peter Littlewood, 
Ben Simons, T.V. Ramakrishnan and Philippe Monthoux for many
helpful discussions.  

\bibliography{UGe2_abbrev,flex_abbrev}

\end{document}